\documentclass[aps,twocolumn,floats,superscriptaddress,floatfix]{revtex4}

\begin{document}

\title{Critical Hysteresis from Random Anisotropy}
\author{Rava A. da Silveira}
\affiliation{Lyman Laboratory of Physics, Harvard University, Cambridge, Massachusetts
02138, U. S. A.}
\author{Stefano Zapperi}
\affiliation{INFM UdR Roma 1 and SMC, Dipartimento di Fisica, Universit\`{a}
\textquotedblleft La Sapienza,\textquotedblright\ P.le A. Moro 2, 00185
Roma, Italy}

\begin{abstract}
Critical hysteresis in ferromagnets is investigated through a $N$-component
spin model with random anisotropies, more prevalent experimentally than the
random fields used in most theoretical studies. Metastability, and the
tensorial nature of anisotropy, dictate its physics. Generically, random
field Ising criticality occurs, but other universality classes exist. In
particular, proximity to $\mathcal{O}(N)$ criticality may explain the
discrepancy between experiment and earlier theories. The uniaxial anisotropy
constant, which can be controlled in magnetostrictive materials by an
applied stress, emerges as a natural tuning parameter.
\end{abstract}

\maketitle

%\section{introduction}
Hysteretic properties of ferromagnetic materials have long fueled applied
research and, more recently, much theoretical interest \cite{Bertotti}. As a
manifestation of the non-equilibrium dynamics of a disordered system with
many degrees of freedom, hysteresis is described naturally in the language
of statistical mechanics. A central aim of theoretical studies of hysteresis
is to elucidate the ways in which microstructural details, such as domain
configurations, lattice structure, impurities or defects, affect macroscopic
properties such as the shape of the hysteresis loop and the Barkhausen noise
statistics.

A non-equilibrium version of the zero temperature random field Ising model
(RFIM) has served to illustrate the competing effects of disorder and
(ferromagnetic) exchange interaction involved in hysteresis \cite{SET-93}.
In three and higher dimensions, at weak disorder the model exhibits a
discontinuous hysteresis loop, which becomes continuous at strong disorder.
These two phases are separated by a critical loop for a given value of the
disorder; as the latter is approached from the weak disorder side, the
macroscopic discontinuity vanishes continuously, resulting in a critical
point characterized by universal scaling laws \cite{SET-93,DAH-96,PER-99}.
The corresponding critical exponents were obtained within a mean field
approximation \cite{SET-93}, perturbatively in a renormalization group
treatment \cite{DAH-96}, and exactly on the Bethe lattice \cite{DHA-97}.
(While the model was originally suggested in part to relate this disorder
induced critical scaling to Barkhausen noise measurements \cite%
{SET-93,DAH-96}, it seems that in most experiments the statistics of the
noise is controlled instead by the depinning transition of domain walls \cite%
{ZAP-98,DUR-00}, which do not emerge simply from an analysis in terms of a
RFIM.)

Direct experimental evidence of disorder induced transitions in ferromagnets
was obtained only recently. A temperature controlled transition was reported
for Co-CoO bilayers \cite{BER-00} and a similar transition was observed in
Gd/W films subjected to different annealing procedures, which induce
variations in the disorder through variations of the grain size \cite%
{BER-00b}. A study of hysteresis loops of Cu-Al-Mn alloys for different Mn
concentrations and temperatures also identified a transition \cite{MAR-03},
and the measured scaling exponents are consistent with those observed for
Co-CoO bilayers. They do not agree, however, with predictions of the RFIM. A
natural explanation proposal for this discrepancy focuses on the nature of
disorder; indeed, while random fields are convenient for theoretical
exploration, they are seldom present in real ferromagnets, which display
more complicated forms of disorder. Prominently, random anisotropies are
present in most ferromagnets, including soft materials, and are believed to
be particularly relevant in amorphous rare earth alloys \cite{ALB-78}.

A disorder induced phase transition was observed numerically in a random 
\textit{infinite} anisotropy model, with exponents close to the RFIM ones 
\cite{VIV-01}, supporting general symmetry arguments that were put forth in
favor of universality \cite{DAH-96}. However, infinite anisotropies pin the
spins to given (random) directions, making each spin Ising-like on its own;
as a result, the model is equivalent to a random field, random bond model.
Furthermore, within a non-equilibrium context symmetry arguments ought to be
taken with a grain of salt; it is known, for example, that the magnetization
may point away from the applied field out of equilibrium, while in
equilibrium minimization of the free energy requires alignment of the two.
Such phenomena are a consequence of the presence of many metastable states
(involved in the dynamics), and more systematic analyses that clarify their
role and substantiate the symmetry arguments are worthwhile. Along these
lines, a renormalization group study of a random field vectorial ($\mathcal{O%
}(N)$) model (RFVM), taking metastability into account, showed that while
one is justified in expecting a critical behavior identical to that of the
RFIM generically, by tuning additional parameters different universality
classes may be visited \cite{DAS-99}.

Here, we analyze a non-equilibrium random anisotropy vectorial model (RAVM),
in which $N$-component spins are subjected to ferromagnetic interactions and
random (finite) anisotropies. The $N=2$ case was proposed in the past as a
model of rare earth alloys, and its hysteretic behavior was studied
numerically \cite{DIE-90}. While in these studies the anisotropy averages to
zero, here we allow for a fixed uniaxial component in addition to a random
background. From the zero temperature spin dynamics, we construct the
appropriate non-equilibrium effective action \cite{DAS-99} which describes
the evolution of the magnet along the hysteresis loop. The most notable
consequence of \textit{metastability} is the generation of a `\textit{random
field} term' in the action; in addition, the latter breaks the rotational
symmetry verified by the \textit{equilibrium} action. As a result, random
anisotropy magnets indeed generically display usual, RFIM exponents (at
least within the domain of validity of the perturbative analysis) for given
values of the applied field and disorder strength. However, there exists a
number of additional universality classes, and in particular a critical
point with $\mathcal{O}(N)$ exponents which is reachable upon tuning of an
additional parameter. The higher (tensorial) nature of the disorder in the
RAVM provides such an additional parameter, namely the uniaxial anisotropy
constant, in a natural fashion. As explained below, proximity to a vectorial
($\mathcal{O}(N)$) critical point may help explain the discrepancy between
experimentally measured exponents and Ising ones.

In the RAVM, $N$-component spins $\vec{s}$ on a $d$ dimensional lattice
interact \textit{via} ferromagnetic nearest neighbor couplings $J_{0}$ and a
spin at site $i$ couples to an anisotropy tensor $K_{i}^{\alpha \beta }$. In
addition, the spins are subjected uniformly to an applied magnetic field $%
\vec{H}$, which varies (adiabatically) in time and hence forces the system
out of equilibrium, through a Hamiltonian 
\begin{equation}
\mathcal{H}=-J_{0}\sum_{<ij>}\vec{s}_{i}\cdot \vec{s}_{j}-\sum_{i}\left( 
\vec{s}_{i}\cdot K_{i}\cdot \vec{s}_{i}+\vec{H}\cdot \vec{s}_{i}\right) ,
\label{eq:H}
\end{equation}%
where $\vec{s}_{i}\cdot K_{i}\cdot \vec{s}_{i}$ stands as a shorthand for $%
\sum_{\alpha ,\beta =1}^{N}s_{i}^{\alpha }K_{i}^{\alpha \beta }s_{i}^{\beta
} $. (Latin indices label lattice sites while Greek indices label spin
components.) The anisotropy tensor may be decomposed into non-random and
random components, and in the simplest (uniaxial) case 
\begin{equation}
K_{i}^{\alpha \beta }=K_{0}n^{\alpha }n^{\beta }+\delta K_{i}^{\alpha \beta
},
\end{equation}%
where $K_{0}$ is the \textit{uniaxial anisotropy constant} and $\hat{n}$ a
unit vector lying along the easy magnetization axis. The random components $%
\delta K^{\alpha \beta }$ are uncorrelated Gaussian random numbers with
vanishing mean and standard deviation $R$, so that the anisotropy tensors
are distributed according to the density%
\begin{equation}
\rho (K^{\alpha \beta })=\frac{1}{\sqrt{2\pi }R}\exp \left( -\frac{%
(K^{\alpha \beta }-K_{0}n^{\alpha }n^{\beta })^{2}}{2R^{2}}\right) .
\end{equation}%
The parameter $R$ plays an analogous role here to that of the width of the
disorder distribution (also called $R$) in random field models \cite%
{DAH-96,DAS-99}. Finally, for the sake of calculational simplicity, instead
of fixed length spins (with, \textit{e.g.}, $\left\vert \vec{s}%
_{i}\right\vert ^{2}=1$ for each site $i$) we consider `soft spins' whose
lengths can take any values. Following Refs. \cite{DAH-96,DAS-99}, for
stability we add to the Hamiltonian a sum of single site terms $\sum_{i}V(%
\vec{s}_{i})$, so that, at each site, a Mexican hat potential $V(\vec{s}%
_{i})=-a\left\vert \vec{s}_{i}\right\vert ^{2}/2+b\left\vert \vec{s}%
_{i}\right\vert ^{4}/4$ prevents the spin from diverging. In the appropriate
limit ($a=b\rightarrow \infty $), soft spins reduce back to unit spins, but
as the length of spins is modified under renormalization, the specific
(bare) values of $a$ and $b$ are irrelevant.

As mentioned, the applied field $\vec{H}$ varies (adiabatically) in time and
forces the spins through a non-equilibrium trajectory. In order to study the
critical behavior of the system, we may confine ourselves to the simple zero
temperature relaxational dynamics%
\begin{equation}
\Gamma \frac{\partial \vec{s}_{i}}{\partial t}=-\frac{\partial \mathcal{H}}{%
\partial \vec{s}_{i}},  \label{eq:dyn}
\end{equation}%
where $\Gamma $ is an effective damping coefficient. We point out, though,
that Eq.~(\ref{eq:dyn}) is certainly not the most realistic choice of
dynamics, which in general includes precession of spins and a more
complicated damping factor, better described by a
Landau-Lifshitz-Gilbert-like equation \cite{Bertotti}. Nevertheless, we
expect the critical behavior not to bear crucially on the specifics of the
microscopic dynamics, and Eq.~(\ref{eq:dyn}) appears as the simplest
candidate for an analytic treatment. In the same vein, in what follows we
make a final, customary simplification in replacing the lattice spins $\vec{s%
}_{i}$ by a continuum vector field $\vec{s}(\vec{x})$.

%\section{The effective action} 

In the continuum approximation, Eq.~(\ref{eq:dyn}) becomes 
\begin{equation}
\Gamma \frac{\partial \vec{s}\left( \mathbf{x},t\right) }{\partial t}%
=J\nabla ^{2}\vec{s}+\vec{H}\left( t\right) +K\left( \mathbf{x}\right) \cdot 
\vec{s}+a\vec{s}-b\left\vert \vec{s}\right\vert ^{2}\vec{s},  \label{eq:dyn2}
\end{equation}%
where the constant $J$ results from the continuum expansion of the exchange
interaction \cite{DAS-99}, $K\left( \mathbf{x}\right) \cdot \vec{s}$ is a
shorthand for the vector field with components $\sum_{\beta =1}^{N}K^{\alpha
\beta }\left( \mathbf{x}\right) s^{\beta }$, and higher orders (in
derivatives and possibly fields) have been neglected. Aspects of the
hysteretic critical behavior are more transparent in the language of a
generating functional \cite{MSR} than directly through the equation of
motion. In the usual fashion \cite{DAH-96,DAS-99}, we introduce an auxiliary
field $\vec{\phi}$ to exponentiate a delta function that forbids any
trajectory that does not obey the equation of motion, resulting, up to some
constant prefactors, in a functional %\begin{widetext} 
\begin{eqnarray}
Z &=&\int \mathcal{D}s\mathcal{D}\phi \exp \left( \int dtd^{d}x\vec{\phi}%
\cdot \left( -\Gamma \frac{\partial \vec{s}}{\partial t}\right. \right. 
\nonumber \\
&&\left. \left. +J\nabla ^{2}\vec{s}+\vec{H}+K\cdot \vec{s}+a\vec{s}%
-b\left\vert \vec{s}\right\vert ^{2}\vec{s}\right) \right)
\end{eqnarray}%
%
%
%
%\end{widetext} 
which captures the possible histories of the system. The advantage of this
procedure is that now one can easily average the generating functional over
the distribution $\rho (K^{\alpha \beta })$ for anisotropy tensors $%
K^{\alpha \beta }$ at all positions, as %\begin{widetext} 
\begin{eqnarray}
\bar{Z} &=&\int \mathcal{D}K^{\alpha \beta }\rho (K^{\alpha \beta })Z 
\nonumber \\
&=&\int \mathcal{D}s\mathcal{D}\phi \exp \left( S_{\text{eff}}\left[ s\left( 
\mathbf{x},t\right) ,\phi \left( \mathbf{x},t\right) \right] \right) ,
\end{eqnarray}%
%
%
%
%\end{widetext} 
with an effective action %\begin{widetext} 
\begin{equation}
\begin{array}{l}
S_{\text{eff}}=\int dtd^{d}x\left\{ \vec{\phi}\cdot \left[ -\Gamma \frac{%
\partial \vec{s}}{\partial t}+J\nabla ^{2}\vec{s}+\vec{H}+K\cdot \vec{s}%
\right. \right. \\ 
\left. \left. +a\vec{s}-b\left\vert \vec{s}\right\vert ^{2}\vec{s}+K_{0}\hat{%
n}\left( \hat{n}\cdot \vec{s}\right) \right] +\frac{R^{2}}{2}\int dt^{\prime
}(\vec{\phi}\cdot \vec{\phi}^{\prime })(\vec{s}\cdot \vec{s}^{\prime
})\right\} ,%
\end{array}
\label{eq:seff}
\end{equation}%
%
%
%
%\end{widetext} 
where $\vec{s},\vec{\phi}$ and $\vec{s}^{\prime },\vec{\phi}^{\prime }$ are
evaluated at times $t$ and $t^{\prime }$, respectively. The effective action
encodes the averaged solutions of Eq. (\ref{eq:dyn2}) and avoids one the
complication of solving a stochastic equation first and then averaging. In
carrying out the average, one trades the stochastic (anisotropy) term with
new terms coupling $\vec{s}$ and $\vec{\phi}$, and the quenched nature of
the disorder is reflected in the presence of a double integral over time
(over $t$ and $t^{\prime }$). In the RAVM, the two terms with $(\vec{\phi}%
\cdot \hat{n})\left( \hat{n}\cdot \vec{s}\right) $ and $(\vec{\phi}\cdot 
\vec{\phi}^{\prime })(\vec{s}\cdot \vec{s}^{\prime })$ replace random field
terms of the form $(\vec{\phi}\cdot \vec{\phi}^{\prime })$ in the RFIM \cite%
{DAH-96} and the RFVM \cite{DAS-99}.

As mentioned, the effective action in Eq. (\ref{eq:seff}) encompasses all
the solutions and as a result is invariant under the transformation $(\vec{H}%
,\vec{s},\vec{\phi})\rightarrow (-\vec{H},-\vec{s},-\vec{\phi})$. The
hysteresis curve, however, is \textit{not} symmetric in general upon
inversion of the magnetic field and magnetization; in particular, the value
of the magnetization at zero field (remanent magnetization), and \textit{%
vice versa} that of the field when the magnetization is zero (coercive
field), do not vanish. This is because the system follows in reality a 
\textit{given} \textit{metastable state} which evolves along with $\vec{H}%
(t) $ (more precisely, each branch of the hysteresis loop corresponds to a
metastable trajectory). Following a trick of Ref. \cite{DAS-99}, we get rid
of the unwanted solutions by shifting the field $\vec{s}$ by a quantity $%
\vec{\sigma}(t)$, which represents the averaged sum of all the `unphysical
minima'; the resulting effective action, $S_{\text{eff, metastable}}$,
encapsulates the magnetization and response function along a branch of the
hysteresis loop. For the sake of simplicity, we consider first the case in
which the field $\vec{H}$ is applied along the easy axis of magnetization, 
\textit{i.e.}, $\vec{H}=H\hat{n}$; we comment below on the general case. We
then expect the magnetization, and hence the vector $\vec{\sigma}$, to lie
along $\hat{n}$ too. Shifting the spin field according to $\vec{s}%
\rightarrow \vec{s}+\sigma (t)\hat{n}$, we obtain %\begin{widetext} 
\begin{equation}
\begin{array}{l}
S_{\text{eff, metastable}}=\int dtd^{d}x\left\{ \phi _{\Vert }\left[ -\Gamma 
\frac{\partial s_{\Vert }}{\partial t}-\Gamma \frac{\partial \sigma }{%
\partial t}\right. \right. \\ 
\left. +J\nabla ^{2}s_{\Vert }+H+\left( s_{\Vert }+\sigma \right) \left(
a-b\left\vert \vec{s}+\vec{\sigma}\right\vert ^{2}+K_{0}\right) \right] \\ 
+\vec{\phi}_{\perp }\cdot \left. \left[ -\Gamma \frac{\partial \vec{s}%
_{\perp }}{\partial t}+J\nabla ^{2}\vec{s}_{\perp }+\vec{s}_{\perp }\left(
a-b\left\vert \vec{s}+\sigma \hat{n}\right\vert ^{2}\right) \right] \right\}
\\ 
+\frac{R^{2}}{2}\int dtdt^{\prime }d^{d}x\left( \phi _{\Vert }\phi _{\Vert
}^{\prime }+\vec{\phi}_{\perp }\cdot \vec{\phi}_{\perp }\;^{\prime }\right)
\\ 
\times \left[ \left( s_{\Vert }+\sigma (t)\right) (s_{\Vert }^{\prime
}+\sigma (t^{\prime }))+\vec{s}_{\perp }\cdot \vec{s}_{\perp }^{\prime }%
\right] ,%
\end{array}
\label{eq:s_shift}
\end{equation}%
%
%
%
%\end{widetext} 
where we have decomposed $\vec{s}=(s_{\Vert },\vec{s}_{\perp })$ and $\vec{%
\phi}=(\phi _{\Vert },\vec{\phi}_{\perp })$ into longitudinal and transverse
components with respect to the direction given by $\hat{n}$.

A number of results may be deduced from the form of the corrected action $S_{%
\text{eff, metastable}}$. In Eq. (\ref{eq:s_shift}), the bare longitudinal
`mass' $a$ (the coefficient of the $\phi _{\Vert }s_{\Vert }$ term) is
dressed into $\tilde{a}_{\Vert }=a-3b\sigma ^{2}+K_{0}$ and the bare field $%
H $ into $\tilde{H}=H-\Gamma \partial _{t}\sigma +\sigma (a-b\sigma
^{2}+K_{0}) $. As, in general, $\tilde{a}_{\Vert }$ and $\tilde{H}$ do not
become small (or vanish) simultaneously at $H=0$, criticality does \textit{%
not} occur at vanishing field. This reflects the non-equilibrium nature of
the trajectory, chosen among many metastable states generated by the
disorder. The more remarkable manifestation of metastability is, however,
the generation of an effective `random field' $\sigma (t)\sigma (t^{\prime })%
\vec{\phi}\cdot \vec{\phi}^{\prime }$ term. As a result, the effective
action in Eq. (\ref{eq:s_shift}) differs from its random field counterpart
by the presence of additional terms of the form $\phi \phi ^{\prime
}s^{(\prime )}$ and $\phi \phi ^{\prime }ss^{\prime }$. Without these terms
the action exhibits a non-trivial (non-Gaussian) critical point below the
upper critical dimension $d_{\text{c}}=6$, which can be characterized by a
perturbative renormalization group treatment in $d=6-\varepsilon $
dimensions \cite{DAH-96,DAS-99}. Now, power counting predicts that this
critical point is stable with respect to the extra random anisotropy terms:
from the natural rescalings $x\rightarrow bx$, $t\rightarrow b^{2}t$, $%
s\rightarrow b^{2-d/2}s $, and $\phi \rightarrow b^{-2-d/2}\phi $, we find a
scaling dimension of $(4-d)/2$ for the $\phi \phi ^{\prime }s^{(\prime )}$
terms and of $4-d$ for the $\phi \phi ^{\prime }ss^{\prime }$ terms, which
are thus \textit{irrelevant} close to $d_{\text{c}}=6$ dimensions.
Consequently, at least within the perturbative domain, criticality in the
RAVM is identical to that in the RFVM \cite{DAS-99} with, generically, RFIM
exponents \cite{DAH-96} reflecting `massless' fluctuations of the
longitudinal component $s_{\Vert }$.

By symmetry, as in the RFVM \cite{DAS-99} there must exist here a $\mathcal{O%
}(N-1)$ critical point representing `massless' fluctuations of the
transverse components $\vec{s}_{\perp }$, corresponding to spontaneous
magnetization in the transverse direction. Then, upon appropriate tuning of
an additional parameter, these two critical points may merge, resulting in a
rotationally invariant vectorial ($\mathcal{O}(N)$) critical point. The
latter occurs for a symmetric action, in particular the effective
longitudinal mass $\tilde{a}_{\Vert }$ and the effective transverse mass $%
\tilde{a}_{\perp }=a-b\sigma ^{2}$ must become small simultaneously. In the
RFVM, simultaneous vanishing of the effective masses and applied field is
possible only if the magnetization vanishes at $H=0$, \textit{i.e.}, for
very `thin' hysteresis loops (with small area) \cite{DAS-99}. Here,
crucially, the higher (tensorial) nature of anisotropy alters this picture:
vector criticality occurs generically at \textit{non-vanishing} values of
the applied field and magnetization. Indeed, since $K_{0}$ modifies both the
field \textit{and} the longitudinal mass, the values of $\tilde{H}$, $\tilde{%
a}_{\Vert }$, and $\tilde{a}_{\perp }$ may become critical simultaneously at
non-vanishing values of $H$ and $\sigma $. As a result, in the RAVM the
hysteresis loop need \textit{not} be `thin' to display vectorial
criticality. The uniaxial anisotropy constant $K_{0}$ may be tuned instead
of the disorder width to reach Ising criticality, or along with the disorder
width to reach vectorial criticality.

We emphasize that this picture is a direct consequence of metastability. By
contrast, the \textit{equilibrium} random anisotropy model \cite{harris1973}
displays a lower critical dimension of $d_{\text{c}}=4$, which is also, if
naive power counting is to be believed, the upper critical dimension.
Systematic studies of the behavior about $d_{\text{c}}=4$ are plagued with a
number of technical difficulties \cite{aharony1975} and, although the weak
disorder phase below $d_{\text{c}}$ was captured in a recent analytical
treatment \cite{feldman}, agreement with experiments \cite{experiments} and
numerics \cite{numerics} is still controversial. While it is certainly
legitimate to ask whether similar difficulties arise out of equilibrium away
from $d_{\text{c}}=6$, we note that, curiously, at least in the perturbative
domain, metastability simplifies the problem.

So far, we have considered the particular case in which the system is
magnetized along its easy axis. A similar analysis may be applied to the
general case, in which the magnetization lies along a direction $\hat{\mu}$
intermediate between those of $\hat{n}$ and $\vec{H}$. As $H$ increases (or
is varied) in time, $\hat{\mu}$ rotates in space. This, however, does not
affect the analysis significantly (calculationally, because different times
effectively decouple in the action). The fact that $\hat{\mu}$ does not lie
along $\hat{n}$ or $\vec{H}$ changes the symmetries, in particular because
additional transverse terms are generated by the shift in $\sigma $. These
seem to allow for the possibility of \ $\mathcal{O}(2)$ and $\mathcal{O}%
(N-2) $ critical points, provided additional parameters may be tuned.

In sum, we have shown that critical hysteresis in the RAVM is described,
generically, by RFIM exponents. Thus, we expect the conclusions of
simulation studies of the random \textit{infinite} anisotropy model \cite%
{VIV-01} to extend to the \textit{finite} anisotropy case in general.
However, we also expect it to be easier to identify vectorial critical
points in the presence of anisotropy than in the presence of random fields,
and a potential explanation for the discrepancy between experiments and
theory lies in a putative proximity of the regime in which experiments are
carried out to such a vectorial critical point. An experimental study in
which various parameters are scanned systematically should reveal whether
proximity to a vectorial critical point is verified. If it is the case, one
expects $\mathcal{O}(N)$ exponents, that cross over to RFIM ones only above
some scale which may be rather large \cite{DAH-96,DAS-99}.

For practical reasons, the standard experimental tuning parameter is
temperature, either annealing or measured temperature, but a number of
interpretation problems are associated with the corresponding techniques. In
the first case, one loses control over changes in microstructure and one is
bound to repeat the experiment on different samples, with possibly large
sample-to-sample variations. In the second case, varying the temperature
modifies simultaneously several physical quantities. In the RAVM, the
natural tuning parameter is the uniaxial anisotropy constant $K_{0}$, rather
than the temperature; it may be used both to reach the RFIM and, along with
a second tuning parameter, to look for vectorial criticality. Tuning $K_{0}$
seems a good experimental possibility for magnetostrictive materials in
which, in the simplest description, an applied stress $\tau $ along the easy
axis shifts the value of the uniaxial anisotropy from $K_{0}$ to $%
K_{0}+3\lambda \tau /2$, where $\lambda $ is the magnetostriction constant 
\cite{Bertotti}.

A more drastic reason for the discrepancy between experiment and theory may
be, of course, that the experimentally predominant form of disorder is
neither of the random field nor of the random anisotropy type. In
particular, in materials used at present in experiments \cite%
{BER-00,BER-00b,MAR-03}, the presence of random bonds and demagnetizing
fields might alter the theoretical picture \cite{ZAP-98}, and so would
putative strong dipolar forces \cite{ZAP-98}. Nevertheless, we expect our
results to be relevant for a wide class of amorphous ferromagnets, and in
particular it would be interesting to check them against experiments on
polycrystals, in which dipolar forces are weak and anisotropy is the
dominant form of disorder.

We thank G. Durin, D.S. Fisher, M. Kardar, and M.A. Mu\~{n}oz for useful
conversations. R.A.S. is grateful for the hospitality at the Statistical
Mechanics of Complexity Center of the INFM, at \textit{La Sapienza}, where
this work was pursued in part. Support from the Harvard Society of Fellows,
the Milton Fund, and the \textit{Fonds national suisse} (R.A.S.) is
acknowledged.

\end{document}